\documentclass[]{spie}  

\usepackage{amsmath,amsfonts,amssymb}
\usepackage{graphicx}
\usepackage[colorlinks=true, allcolors=blue]{hyperref}
\usepackage{subfig}
\usepackage{siunitx}

\title{Ion Irradiation of GridPix Detector at Bonn Isochronous Cyclotron}

\author[a]{Vladislavs Plesanovs}
\author[c]{Hemanth Manikantan}
\author[b]{Dennis Sauerland}
\author[c]{Paolo Soffitta}
\author[c]{Carlo Lefevre}
\author[c]{Ettore Del Monte}
\author[a]{Jochen Kaminski}
\author[b]{Reinhard Beck}
\author[a]{Klaus Desch}

\affil[a]{Rheinische Friedrich-Wilhelms-Universität Bonn, Physics Institute, FTD, Kreuzbergweg 24, Bonn, Germany}
\affil[b]{Rheinische Friedrich-Wilhelms-Universität Bonn, Physics Institute, HISKP, Nussallee 14-16, Bonn, Germany}
\affil[c]{Insituto di Astrofisica e Planetologia Spaziali, Via del Fosso del Cavaliere 100, 00133, Roma}

\begin{document} 
\maketitle

\begin{abstract}

The potential successor of the IXPE --- the EXPO or Enhanced X-ray Polarimetry Observatory will have a new X-ray polarimeter based on GridPix technology. To certify this detector for operation within the expected radiation environment conditions, the irradiation tests were performed at the Bonn Isochronous Cyclotron at HISKPabbreviation\footnote{ stands for Helmholtz-Institut für Strahlen- und Kernphysik or Helmholz Institute for Radiation and Nuclear Physics}, University of Bonn. Two beam modes were used --- proton beam at 14\,MeV and 100\,MeV $^{14}\mathrm{N}^{5+}$ ions. The aforementioned GridPix detector contains the readout ASIC, Timepix, with $256\times256$ pixel grid with a pitch of \SI{55}{\micro\meter} and covered with a protection layer. The Al grid stands on the SU8 pillars above pixels forming the amplification stage. The alignment of grid holes and silicon pixels enable primary electron detection. This paper focuses on the description and results of these testbeams. The accumulated ion rates exceed by far the expected limits without a significant deterioration in chip performance.

\end{abstract}


\section{INTRODUCTION}
\label{sec:intro}  

The new generation of X-ray optics for X-ray polarietry requires ASICs and detectors with significantly higher count-rate capabilities and much shorter dead times than those employed in IXPE. The GridPix detector \cite{KOPPERT-gridpix-sem}, proposed for the Enhanced X-ray Polarimetry Observatory (EXPO \cite{Soffitta2026}) mission, is a promising candidate to meet these requirements. To assess its suitability for space applications, we adopted the same qualification strategy previously followed for the Gas Pixel Detector, which underwent irradiation tests with 500~MeV\,nucleon$^{-1}$  ions about twenty years ago \cite{Bellazzini2010}, with even a larger deposited energy density. 

To ensure the survivability of the detector in space, on-ground qualifications for radiation tolerance (total ionizing dose) along with the vulnerability to single-event effects (SEU and SEL) must be evaluated. Furthermore, gas detectors using multiplication stages such as the GridPix require evaluation for robustness against heavy ion-induced sparks. The extensive and successful use of Timepix as a space-borne radiation detector \cite{kroupa2015semiconductor,gohl2019study,granja2026miniaturized,furnell2019first} demonstrates that SEU and SEL effects do not pose a significant risk to its operation in orbit. Motivated by this, in this work, we focus on exposing the detector to ionizing radiation to evaluate its tolerance to the total ionizing dose and its survivability against heavy-ion-induced sparks. 

The major cause of these is ionizing charged particles in space, namely trapped particles in the Van Allen radiation belts and Galactic Cosmic Rays (GCRs). Among these, Fe ions in GCRs are considered the most significant, as they deposit substantial energy in the detector and have a comparatively high flux among heavy ions. We computed the energy deposited by Galactic Cosmic Ray Fe ions in a gas cell filled with an Ar/CO$_2$ 70/30 gas mixture at 1 bar pressure, in a low Earth orbit of 600 km, 30$^\circ$ inclination over 20 years, using SPENVIS\footnote{\url{https://www.spenvis.oma.be/}} and SRIM-TRIM\footnote{\url{http://www.srim.org/}}, which yields a value of $\sim86$ GeV \cite{manikantan2025legacy}. This value was used as the target radiation dose for irradiating the GridPix detector.

Using Geant4 simulations, we estimated the energy deposited in the GridPix detector's active gas volume by 97.5 MeV $^{14}\mathrm{N}^{5+}$ ions (7 MeV nuc$^{-1}$), the ion species available at the beamline, shot into the active volume through two layers of aluminum (20 and 25 $\mu$m thick) at the exit of the beamline, and a 2 $\mu$m Mylar window serving as the detector's X-ray window. The mean energy deposited by 97.5 MeV $^{14}\mathrm{N}^{5+}$ ions in the active region was computed to be 13 MeV. Therefore, with a total of $\sim7000$ ions incident on the active volume, we achieve the total radiation dose from 20 years in LEO. For a possible use at the Sun-Earth Lagrangian L2 point, the charged particle rate is expected to be up to $\sim20\times$ larger than in LEO, estimated by comparing the rates reported for Suzaku/XIS in LEO \cite{Yamaguchi2006} and SRG/eROSITA at L2 \cite{Perinati2024}. Taking this factor into account, we decided on a target ion count of $\sim150,000$.

\begin{figure}[]
    \centering
    \subfloat[\label{fig:geometry}]{
    \includegraphics[width=0.35\linewidth]{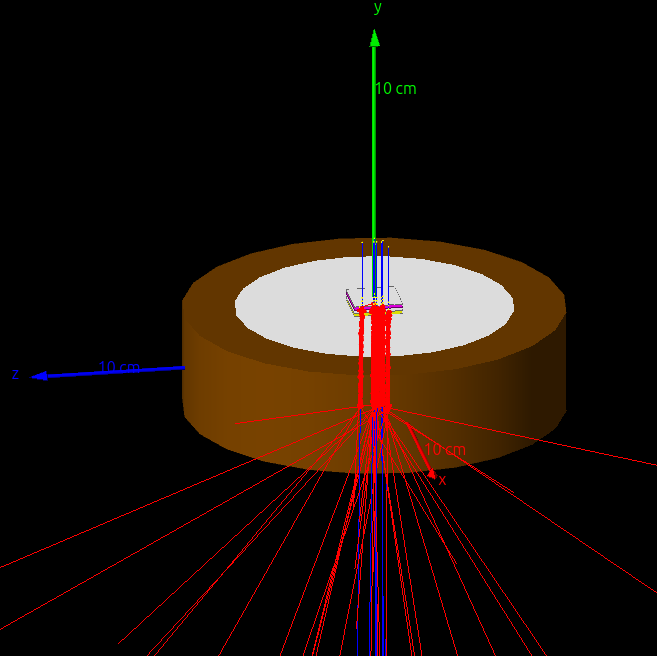}}
    \subfloat[\label{fig:edump}]{
    \includegraphics[width=0.5\linewidth]{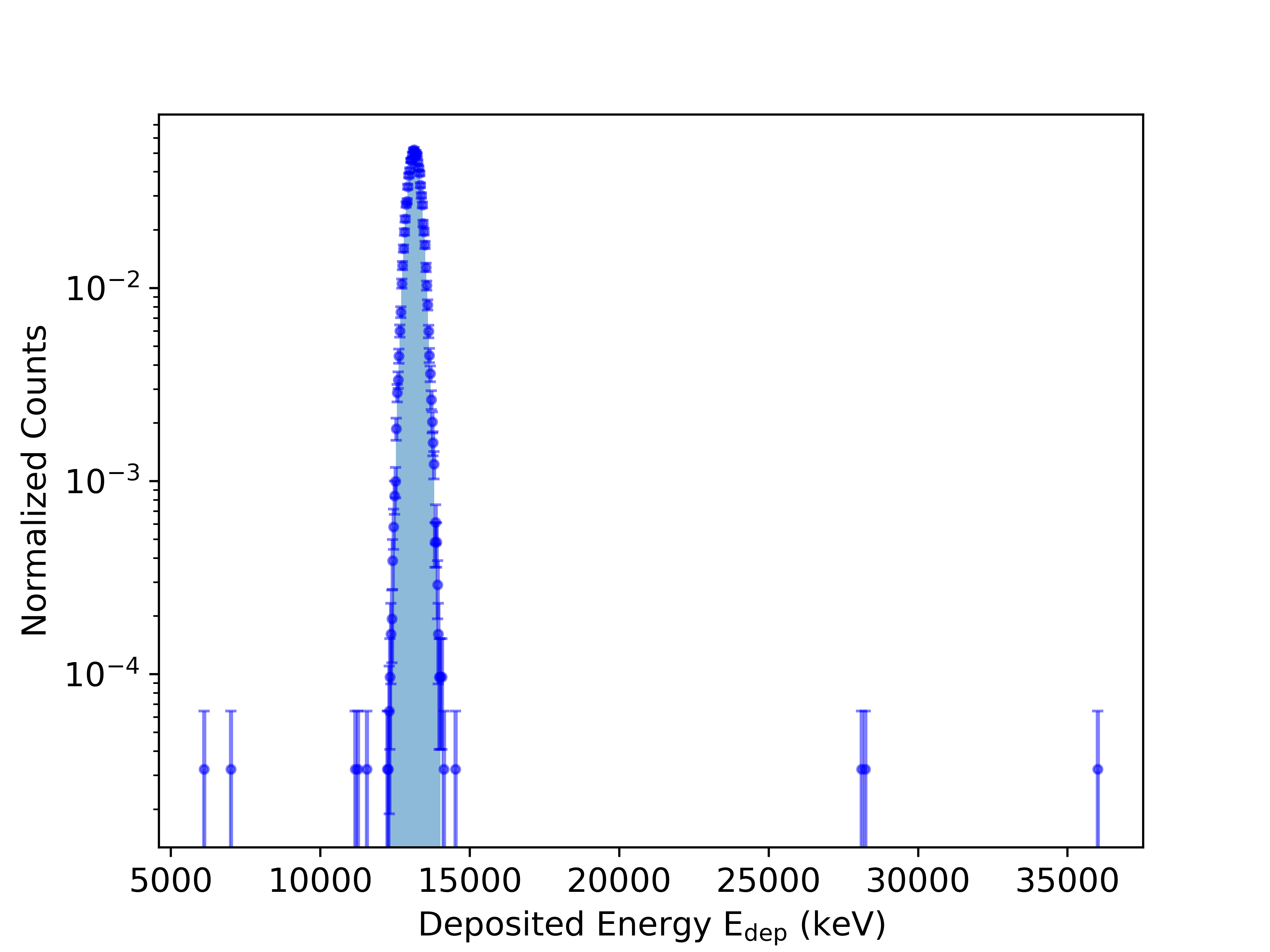}}
    \caption{Geant4 simulation of 7 MeV nuc$^{-1}$ $^{14}\mathrm{N}^{5+}$ ions incident on the GridPix active volume through two layers of aluminum and a Mylar window. The mean energy deposited in the active volume is estimated to be 13 MeV per ion.  Figure~\ref{fig:geometry} shows the simulated geometry of the detector, featuring a cylindrical gas cell (white) surrounded by a PMMA body (brown), and Figure~\ref{fig:edump} shows the resulting energy deposit histogram.}
    \label{fig:simulations}
\end{figure}

\section{Cyclotron Facility}
\label{sec:cyclotron}

\begin{figure}[htbp]
    \centering
    \subfloat[\label{fig:cyclotron-ecr}]{
    \includegraphics[width=0.35\linewidth]{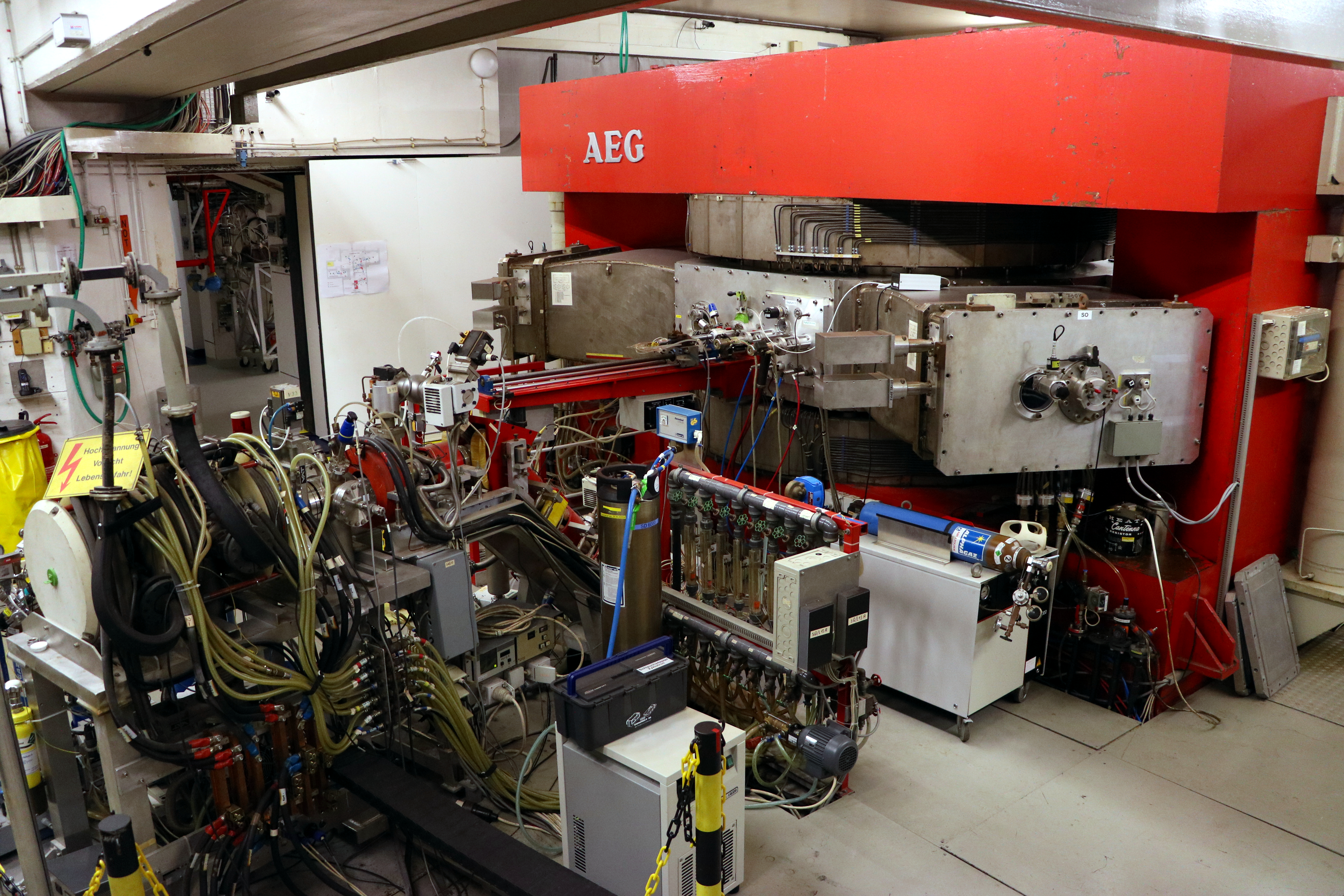}}
    \subfloat[\label{fig:cyclotron-schema}]{
    \includegraphics[width=0.33\linewidth]{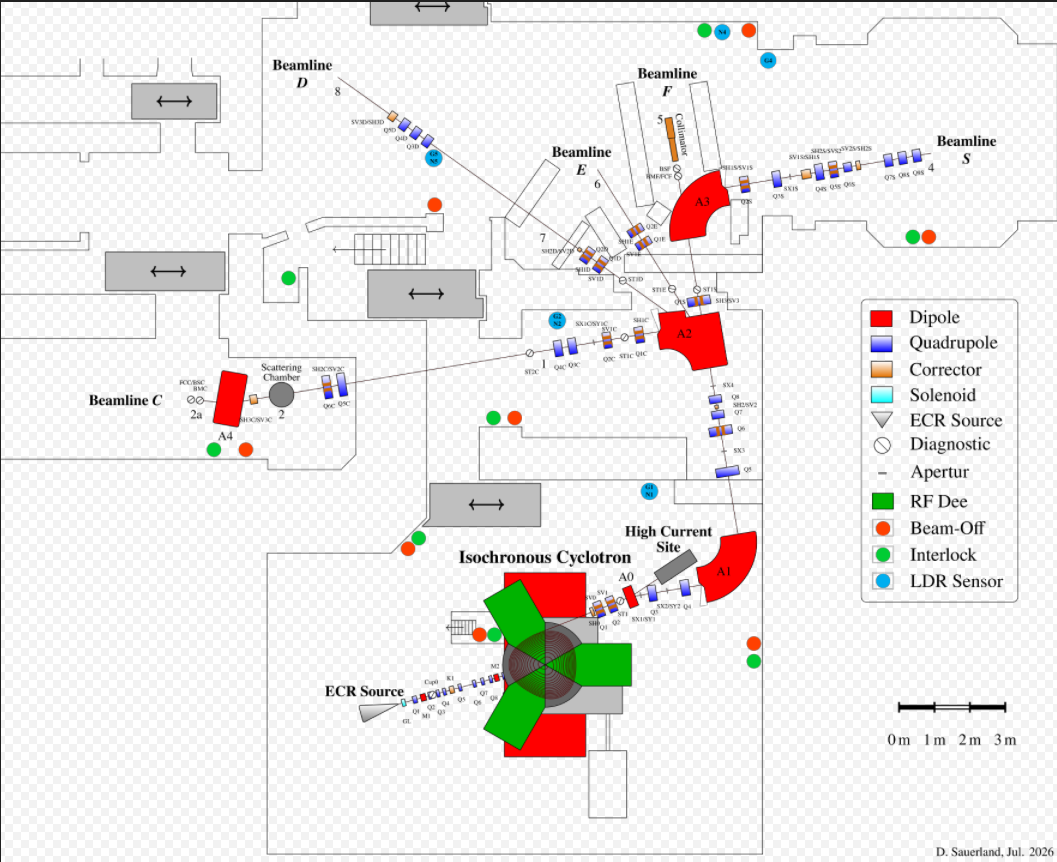}}
    \caption{Photograph of the Bonn Isochronous cyclotron (a) and the layout of the connected beamlines and extraction points (b).}
    \label{fig:cyclotron}
\end{figure}

The accelerator facility features the Bonn Isochronous Cyclotron\footnote{\url{https://www.zyklotron.hiskp.uni-bonn.de/}} (figure \ref{fig:cyclotron-ecr}) with several beamlines and extraction points for various experimental purposes (figure \ref{fig:cyclotron-schema}). The cyclotron is capable to produce charged particle beams with typical kinetic energies of 7--14\,MeV per nucleon. The two external Electron Cyclotron Resonance (ECR) ion sources are used to produce various types of beams ranging from protons to $^{16}\mathrm{O}^{6+}$. For protons and deutrons the beam can be polarized.

The beamline C is used as a test site for silicon-based detector irradiation with a remotely-controlled precision table and offers all required infrastructure for GridPix irradiation studies. The characteristic beam currents at which the cyclotron operates range from \SI{2}{\micro\ampere} down to \SI{20}{\nano\ampere}, which is limited by the capabilities of beam monitoring devices. These currents translate to particle rates in order from GHz to MHz, which are far too high to reproduce the particle fluence expected during a 20-year mission in LEO or at L2. 

Thus a series of preparatory measurements and cyclotron runs were conducted, with the last round using GridPix detector in a parallel beam mode at 14\,MeV energy to verify that the chosen method for beam current attenuation achieves significantly lower rates of irradiation. Only then the final series of measurements was conducted with $^{14}\mathrm{N}^{5+}$ ions at $\mathrm{E}_{\mathrm{kin.}}$=100\,MeV energies. 

\section{GridPix Detector}
\label{sec:gridpix}

\begin{figure}[htbp]
    \centering
    \subfloat[\label{fig:grid-upclose}]{\includegraphics[width=0.32\textwidth]{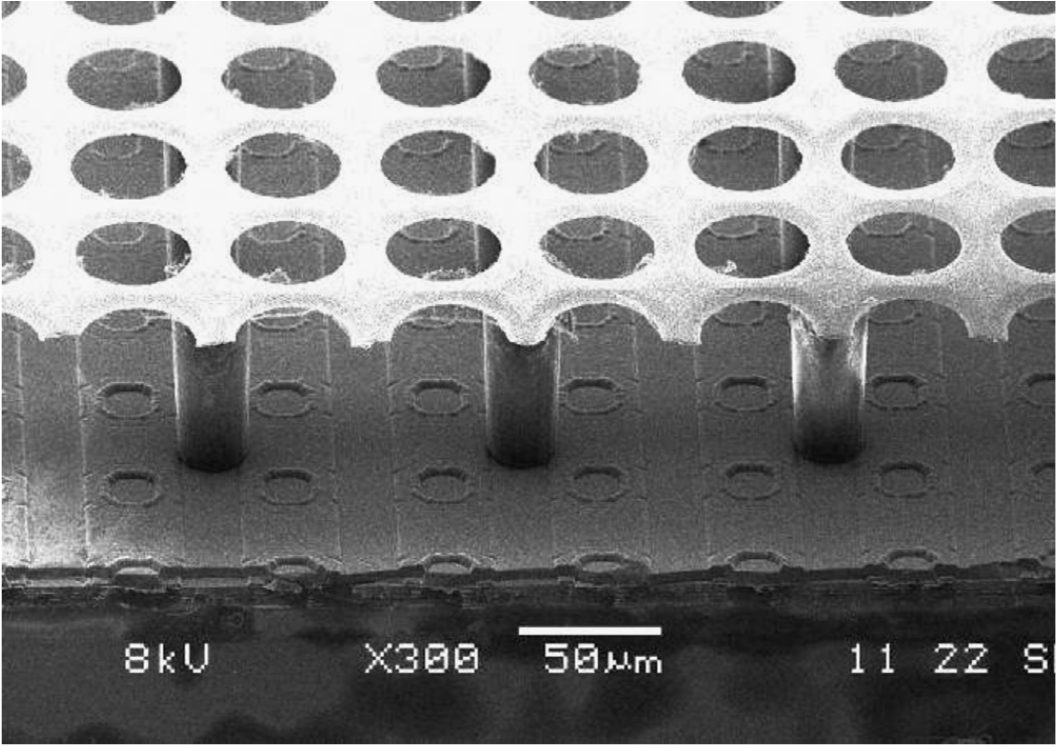}} 
    \subfloat[\label{fig:gridpix-carrier}]{\includegraphics[width=0.30\textwidth]{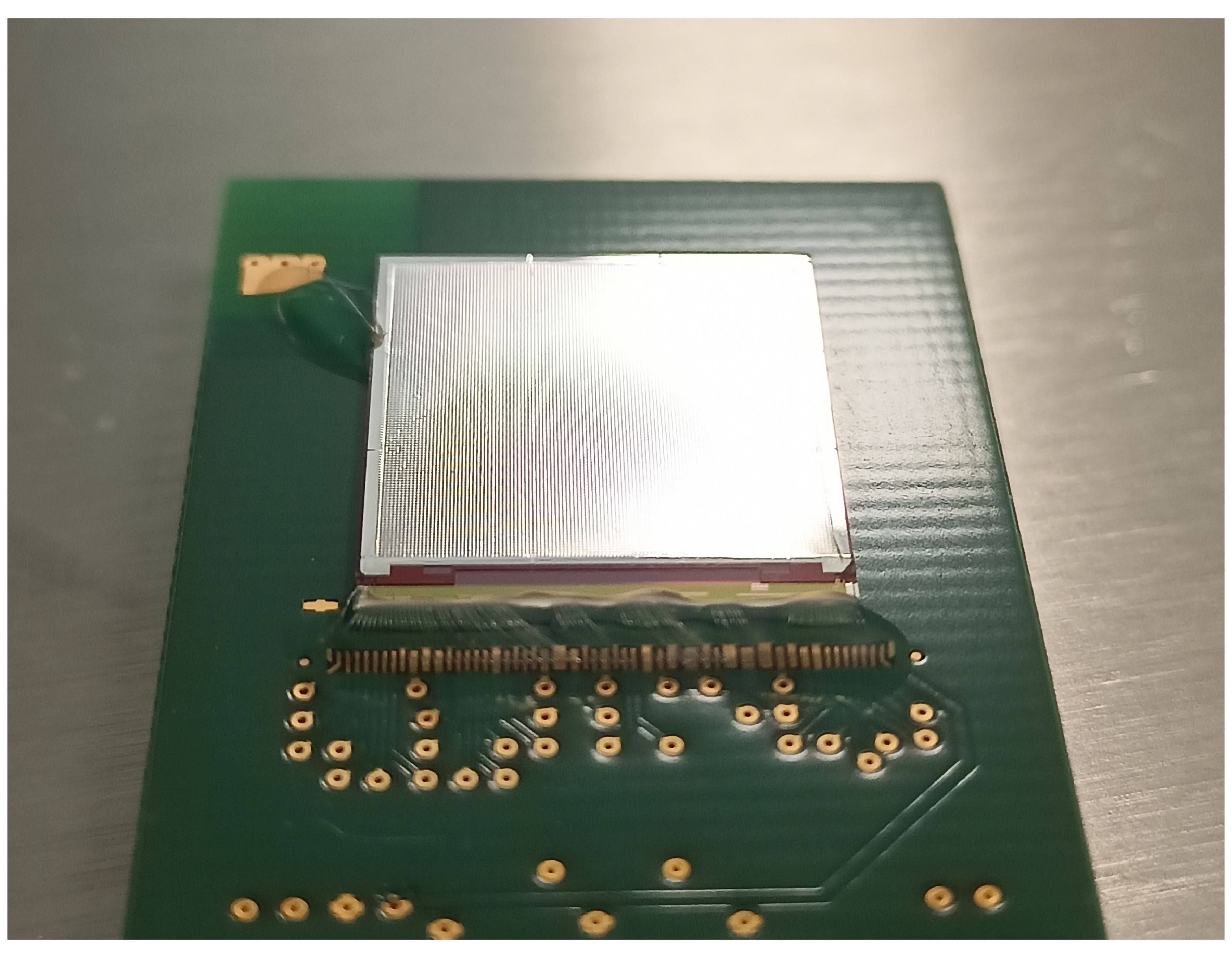}}
    \subfloat[\label{fig:detector-expl}]{\includegraphics[width=0.15\textwidth]{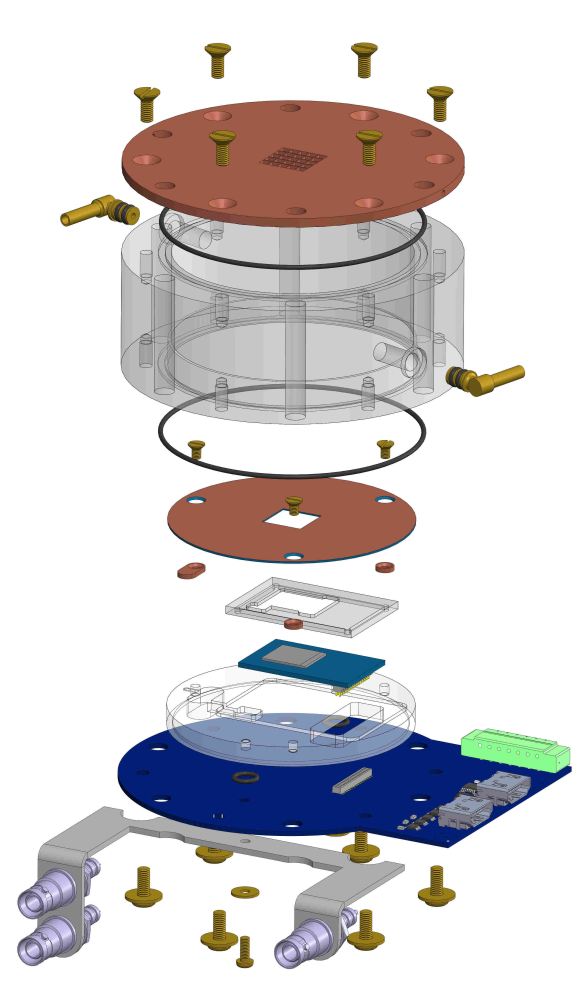}}
    \caption{Constituents and assembly of the GridPix detector. The scanning electron microscope picture of the Integrated Aluminum grid is presented in (a)\cite{KOPPERT-gridpix-sem}. Photograph of the GridPix glued on top of the carrier PCB in the cleanroom environment (b) with visible data, power and HV wire-bonds. The expanded CAD model of the mechanical assembly of the GridPix time projection chamber (c)\cite{ToSchi-thesis}.}
    \label{fig:gridpix}
\end{figure}

The GridPix\cite{KOPPERT-gridpix-sem} detector utilizes integrated grid microstructure on a silicon pixel readout chip. In particular, the \SI{2}{\micro\meter} thick aluminum grid with holes is produced above the surface of the readout ASIC and stands on \SI{50}{\micro\meter} pillars of SU-8 photoresist as shown in figure \ref{fig:grid-upclose}. The photolitographic fabrication process allows to precisely align the grid holes with the readout pixels beneath and enable detection of single primary electrons.

The readout chip in the GridPix assembly is the Timepix \cite{Timepix} and its surface is covered by 4--\SI{8}{\micro\meter} thick silicon nitride layer which acts as a spark protection. The chip's readout matrix features 256$\times$256 pixel grid with a pitch of \SI{55}{\micro\meter}. The Timepix allows to record the events in either Time-over-Threshold (ToT) or Time-of-Arrival (ToA) modes (but not together) in a frame-based readout with a dead-time of 25\,ms.

The GridPix is glued and wire-bonded on a carrier PCB. In figure \ref{fig:gridpix-carrier} one can see the the chip used in this project. In the greater scheme of the gas TPC this carrier PCB is situated on the intermediate PCB and a PMMA stand to stabilize the PCBs. Next is the anode PCB with copper layer, which is placed at an offset of 1\,mm above the surface of the aluminum grid. The active gas volume is enclosed by PMMA cylinder with a copper cathode plate 3\,cm above the anode. In a typical operation mode the cathode, anode, and grid are set to negative potential with the protection layer of the GridPix being the ground. 

In a typical event, a charged particle enters the gas medium and leaves primary ionization pairs along its track. The free electrons drift towards the GridPix, where they are multiplied between the aluminum grid and readout matrix surface leaving high-resolution image. 

\section{Irradiation Results}
\label{sec:Irradiation}

\subsection{Proton Beam}
\label{sec:proton-beam}
The aim of the preparatory measurements with the proton beam was to verify that the beam scraping reduced the beam rate down to the range compatible with the goals of this measurement campaign. The beam optics upstream of beamline C feature a vertical and horizontal scrapers SX1C and SY1C (left of the dipole magnet A2 in figure \ref{fig:cyclotron-schema}). During the measurements with a proton beam the SY1C was fixed to an opening of 0.2\,mm while the SX1C was varied. 

In this beam mode the GridPix was equipped with a PMMA cylinder with a circular opening of 8\,mm in diameter and 1.5\,cm height above the anode level. The hole was covered by \SI{50}{\micro\meter} Kapton foil to minimize energy loss of protons during entry into the gas volume (typical thickness of PMMA wall is 40\,mm). This was sufficient measure since the detector was operated at a slight overpressure of 1.025\,atm.

\begin{figure}[htbp]
    \centering
    \subfloat[\label{fig:proton-tracks}]{
    \includegraphics[width=0.31\textwidth]{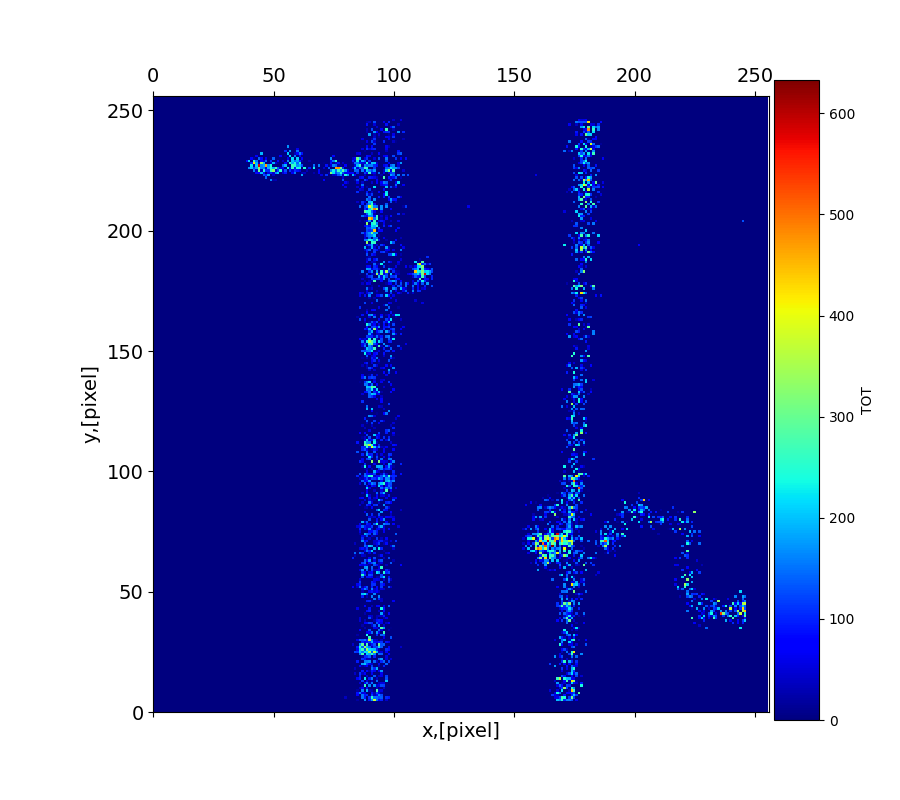}}
    \subfloat[\label{fig:scraper-rate}]{
    \includegraphics[width=0.28\textwidth]{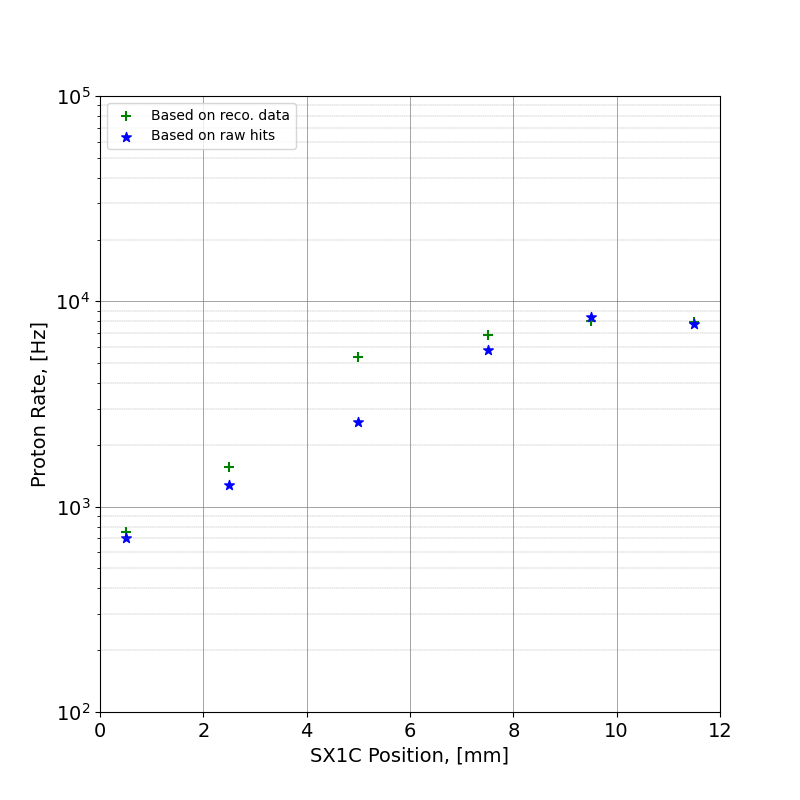}}
    \caption{An example of proton events at $\mathrm{E}_{\mathrm{kin.}}=14$\,MeV recorded in ToT mode (a) and rate of the proton beam versus the horizontal scraper SX1C opening in mm units (b).}
    \label{fig:proton-figs}
\end{figure}

The main results of the proton beam runs are presented in figure \ref{fig:proton-figs}. The figure \ref{fig:proton-tracks} shows well-defined proton tracks recorded at narrow scraper openings. Two individual protons and the associated delta-electron branches are clearly visible.

The figure \ref{fig:scraper-rate} shows the desired confirmation that the scraping of the proton beam worked as expected. Rates on the order of 100\,Hz were achieved and a clear trend of rate increase with increasing SX1C opening is present. The seeming saturation at higher scraper settings is a result of the zero-suppressed readout mode of the readout software. Each track in figure \ref{fig:proton-tracks} produces 1000--1500 pixels hits on average. In the aforementioned readout software mode the maximum number of pixel hits per frame is 4096, which are exceeded by proton tracks at high SX1C settings. Nevertheless, the figure \ref{fig:scraper-rate} shows that the beam rates can be significantly brought down and ensure that the GridPix can be exposed to an ion beam in a non-destructive measurement.

\subsection{Nitrogen Beam}
\label{sec:nitro-beam}

The final irradiation campaign investigated the response of the  GridPix detector to $^{14}\mathrm{N}^{5+}$ ions at 100\,MeV kinetic energy. During these measurements the GridPix detector was mounted directly to the beamline vacuum system at the extraction point. IN this configuration the ions were piercing the GridPix perpendicular to its readout plane. 

Figure \ref{fig:ion-frames} presents typical nitrogen ion events recorded in both ToT ({\ref{fig:ion-frame-tot-4ions}, \ref{fig:ion-frame-tot-3ions-spark}) and ToA (\ref{fig:ion-frame-toa}) modes. These frames show multiple ion tracks accompanied by the delta-electron branches. One of the ions produces a spark visible as a vertical column of responding pixels. Such discharges lead to a spike in the power consumptions of the Timepix chip, causing a shift of the thresholds in corresponding columns of the pixel matrix and resulting in characteristic vertical artifacts (as seen in figure \ref{fig:ion-frame-tot-3ions-spark}). The estimated rate of the ion-induced sparks tends to be $\approx2$\% at the highest amplification voltages. The figure \ref{fig:ion-frame-toa} shows distinct four ion events with the arrival times of the corresponding hits color-coded in the range from 0--11810 digital counts, corresponding to a time window of \SI{287.5}{\micro\second}.

\begin{figure}[htbp]
    \centering
    \subfloat[\label{fig:ion-frame-tot-4ions} 4 ions, ToT mode]{
    \includegraphics[width=0.31\textwidth]{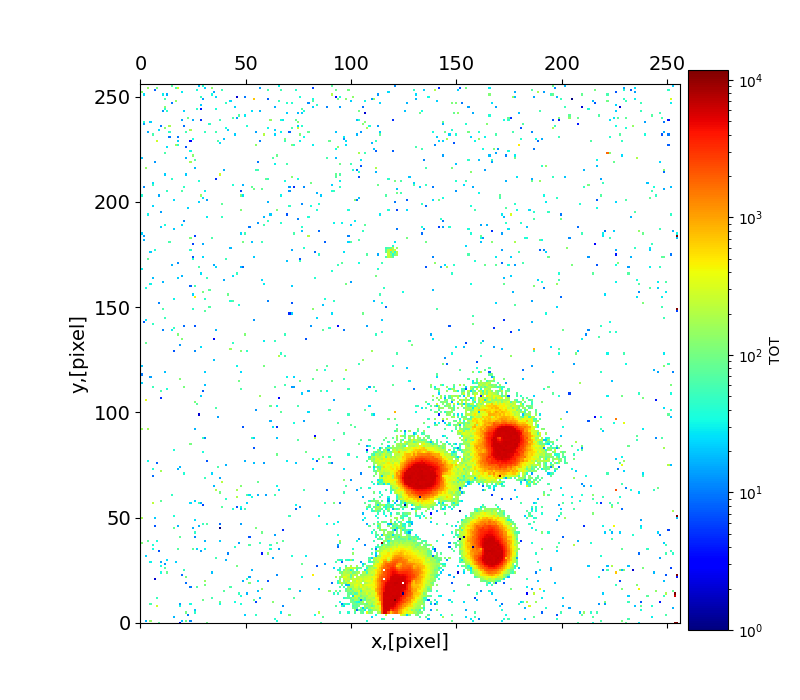}}
    \hspace{0.01\textwidth}
    \subfloat[\label{fig:ion-frame-tot-3ions-spark} 3 ions with spark, ToT mode]{
    \includegraphics[width=0.31\textwidth]{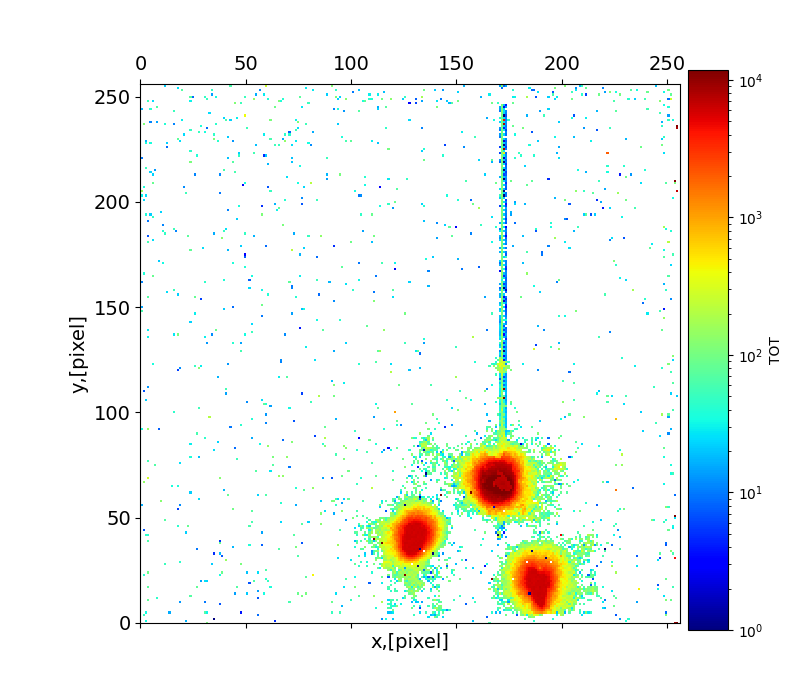}}
    \hspace{0.01\textwidth}
    \subfloat[\label{fig:ion-frame-toa} 4 ions, ToA mode]{
    \includegraphics[width=0.31\textwidth]{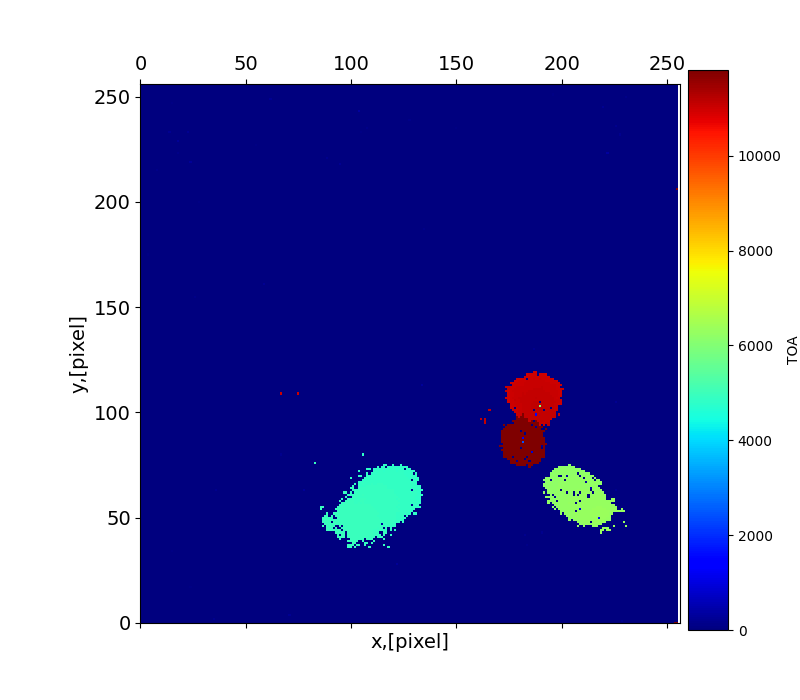}}
    \caption{Examples of the recorded frames with several $^{14}\mathrm{N}^{5++}$ ions in the ToT and ToA counting modes.}
    \label{fig:ion-frames}
\end{figure}

\begin{figure}[htbp]
    \centering
    \subfloat[\label{fig:fe-spectrum}]{\includegraphics[width=0.4\textwidth]{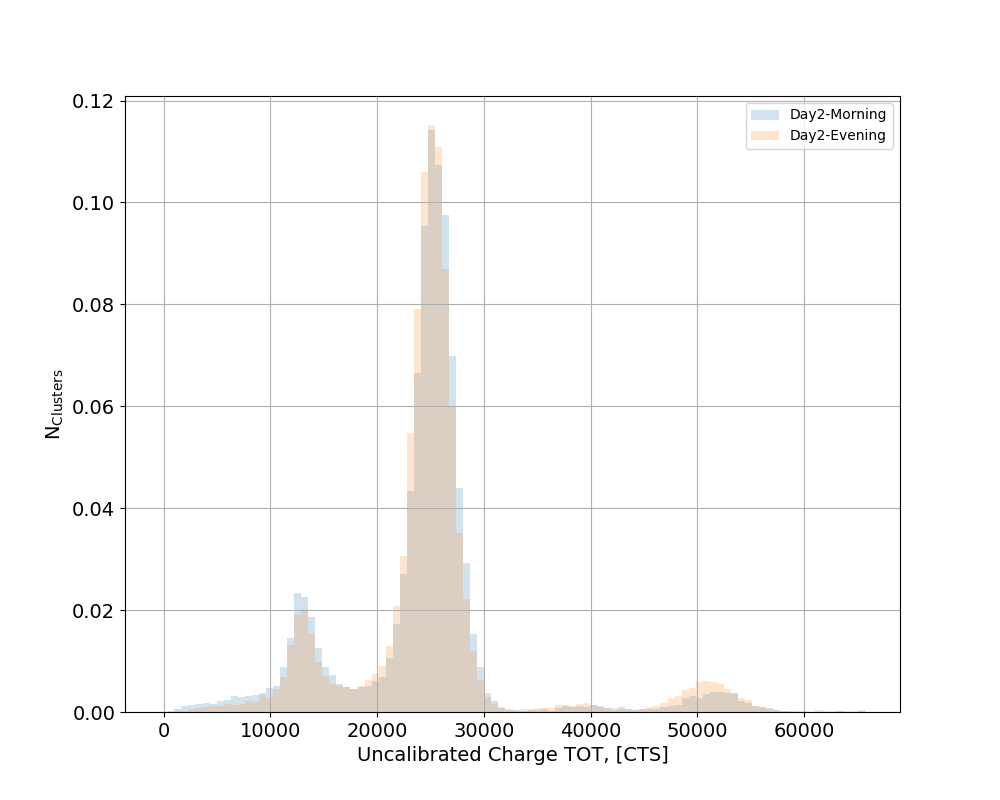}}
    \hspace{0.01\textwidth}
    \subfloat[\label{fig:pixel-diff}]{\includegraphics[width=0.4\textwidth]{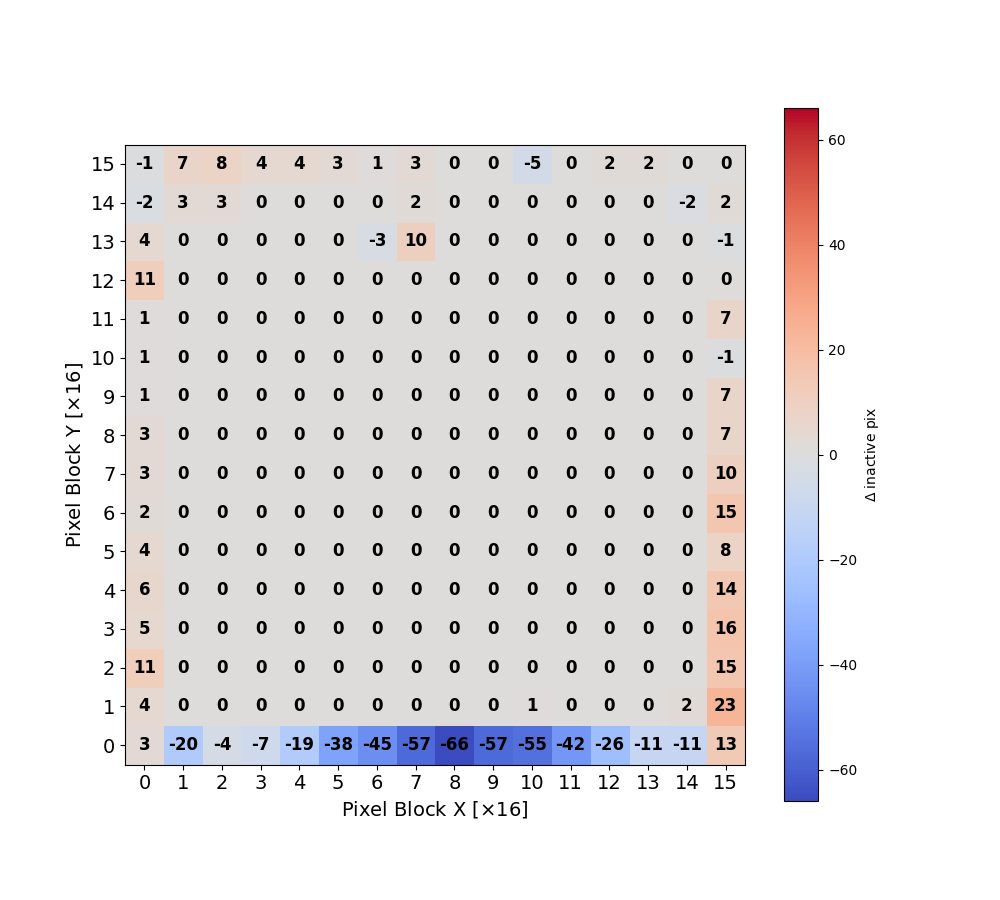}}
    \caption{\label{fe-spectrum-and-pix-diff}
    Spectra of uncalibrated charge in ToT counts for $^{55}\mathrm{Fe}$ runs before and after ion irradiation on one of the days of the beamtime (a). Map of the difference in pixel responsiveness segmented in 16$\times$16 pixel blocks (b). The positive number indicates increase in number of pixels that did not respond in the measurement after GridPix exposure to ion beam, while negative number indicates increase in the number of responding pixels.  
    }
    \label{fig:placeholder}
\end{figure}
To monitor GridPix performance a set of measurements were performed using an X-ray source ($^{55}\mathrm{Fe}$). Figure \ref{fig:fe-spectrum} compares the uncalibrated charge spectra recorded by GridPix before and after irradiation on day 2 of measurements. Both, the escape and 5.9\,keV peaks are visible\footnote{The higher peaks correspond to events where two or more X-ray clusters overlap and can not be separated} with no significant shift between them. 

Potential ion-induced degradation of the pixel matrix was evaluated by comparing two $^{55}\mathrm{Fe}$ runs on the basis of normalized hit rate per pixel. The figure \ref{fig:pixel-diff} shows the accounted difference in the absolute number of pixels in 16$\times$16 pixel blocks of the readout matrix. The relatively large variation of values along the outer perimeter of the matrix is dictated by the edge effects of the electric field. More important is the absence of significant change in the inner area of the pixel matrix. Although the beam was mainly concentrated primarily in the lower half of the active surface, only a few pixels did not respond in the second measurement with $^{55}\mathrm{Fe}$ source. The visible changes in the upper half of the matrix correspond to a pre-existing defects rather than the ion-induced damage.

\begin{figure}
    \centering
    \includegraphics[width=0.5\linewidth]{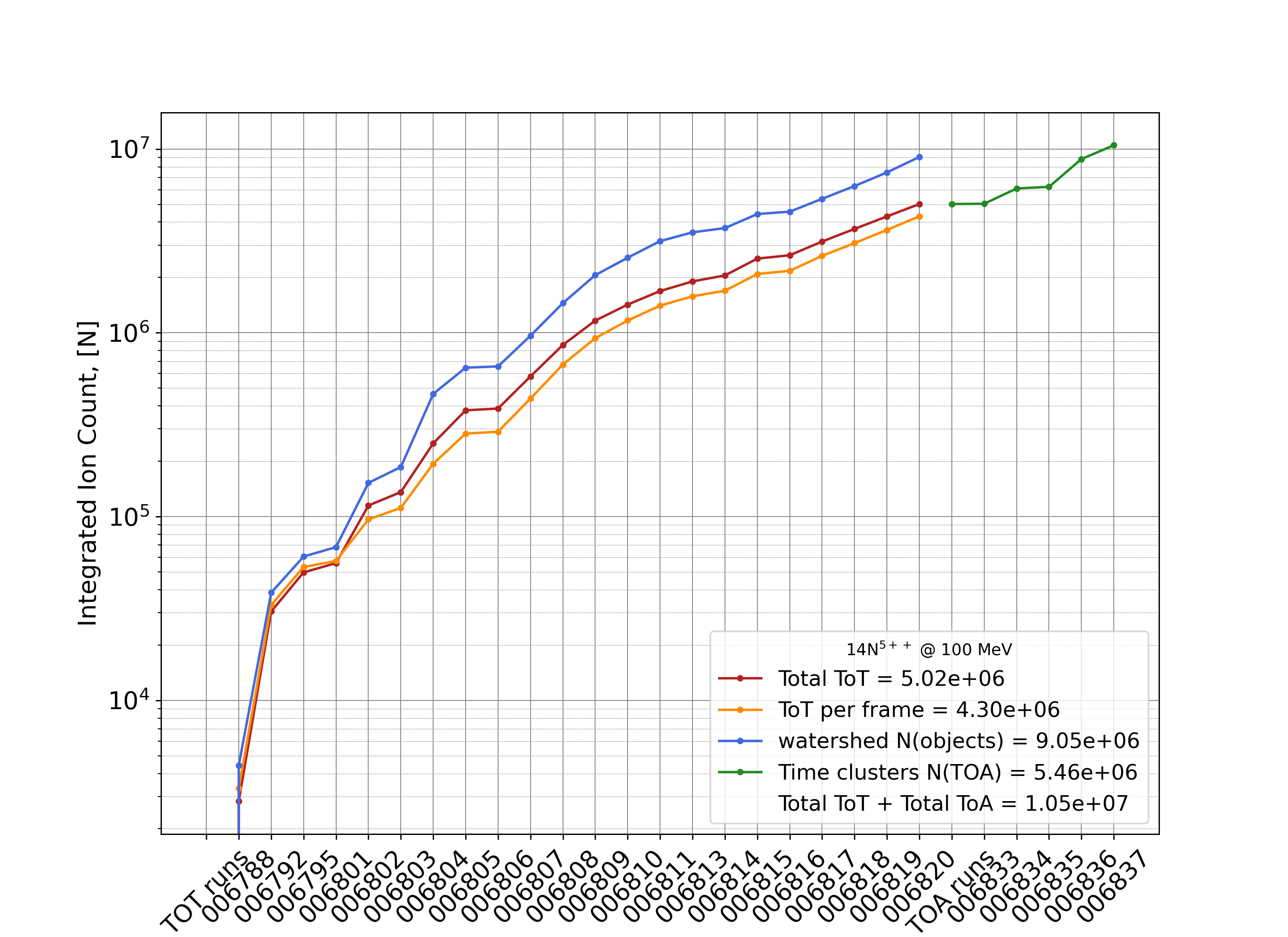}
    \caption{Integrated number of nitrogen ions $^{14}\mathrm{N}^{5++}$ at the energy of 100\,MeV over the runtime of two days. The orange line represents the counting results based on total ToT count division in a frame over the average ToT counts for single ion clusters. Red line represents the total ToT counts in a dataset divided by the average ToT counts per single ion. The blue line represents "watershed" clustering. The green line shows the accumulated ion count in Timepix ToA mode on day 2.}
    \label{fig:ion-total-number}
\end{figure}

The conservative counting of the total number of nitrogen ions that passed through the GridPix during the beamtime is presented in figure \ref{fig:ion-total-number}. Two days of measurements are separated according to the acquisition mode on x-axis with ToT mode runs from first day and ToA data from second day. Four curves show integrated ion count based on four distinct methods. The bottom curves (orange and red) reflect counting of equivalent charge. Single ion clusters were identified and the corresponding average charge per-cluster was used to count ions on the whole data set level or per-frame basis. The blue line represents "watershed"\cite{Beucher2000-watershed} clustering results. In this approach the binary hit map is converted to Euclidean distance transform. Resulting local maxima define cluster seed markers. Then the watershed transform is applied to the inverted distance map to separate overlapping cluster regions and prevent merging of neighboring clusters.

Charge-based ion counting is subject to several systematic effects. Firstly, it does not consider  tracks' inclination. The inclined ion tracks create more hits with ToT counts per track in compared to perpendicular (straight) ion tracks. For a straight ion track a significant part of the primary electrons end up in the same pixel holes leading to multiple consecutive amplification events above a single pixel within a short time interval. Resulting reduction of gas gain leads to decrease in measured charge per pixel. 

The second major effect comes from events with high cluster multiplicity in a single frame in ToT mode. In the Timepix chip, after a pixel detects a signal it processes the hit and remains inactive until the frame is read out. Consequently, overlapping ion clusters share pixel regions where charge information from one of the ions is partially lost. As a result, in counting process the average charge per ion cluster is slightly underestimated leading to underestimation of number of ions in the data set.  

The ion counting for Timepix ToA mode relies on clustering hits in time bins, which works well within \SI{287.5}{\nano\second} window. However, many frames contain multiple ion events recorded after the overflow of the ToA counter, causing them to be assigned to the same time bin. Thus the total number of recorded clusters also underestimates the actual number of ions traversing GridPix detector.

Considering these limitations, all the counting results in figure \ref{fig:ion-total-number} show that the ion dose inflicted on the GridPix has exceeded the limit of qualification for L2 orbit by the end of the first measurement day. The total dose of at least $\mathrm{10}^{7}$ ions was collected over the full beam time.

\section{Conclusions}
The measurements conducted show that GridPix operated safely and reliably and survived both types of particle beams without a significant impact on its performance. The spark protection, when they occasionally occurred (< 2\% of the Nitrogen ions rate), shows that the expected performance as the detector is safely recovered. As a result, conservative estimates show that the detected particle numbers, at least 10$^{7}$, are well above the limits (1.5 10$^{5}$) for qualification to both the LEO and L2/L1 orbits.

\acknowledgments 
This work was supported in part by the
Italian Ministry of Foreign Affairs and International Cooperation

%

\bibliography{report} 
\bibliographystyle{spiebib} 

\end{document}